\documentclass{optica-article}

\journal{opticajournal} 

\articletype{Research Article}

\usepackage{numprint}
\usepackage{siunitx}
\usepackage{amsmath}
\usepackage{mathtools}

\renewcommand{\i}{\mathrm{i}}
\newcommand{\e}{\mathrm{e}}
\renewcommand{\d}{\mathrm{d}}
\renewcommand{\vec}[1]{\mathbf{#1}}
\newcommand{\vect}[1]{\mathbf{#1}}
\DeclareMathOperator*{\argmin}{arg\,min}

\begin{document}

\title{Digital confocal imaging in holographic optical coherence tomography at high space-bandwidth product}

\author{N.~P.~Klooster,\authormark{1} S.~A.~Thakur,\authormark{1} F.~Zeisberger,\authormark{1} J.~R.~Hagen,\authormark{1} S.~Eggert,\authormark{1} U.~Najar,\authormark{1} and D.~Hillmann\authormark{1,2,*}}

\address{\authormark{1}Department of Physics, Vrije Universiteit Amsterdam, 1081 HV Amsterdam,
Netherlands\\
\authormark{2}Faculty of Physics, Bielefeld University, 33615 Bielefeld, Germany}

\email{\authormark{*}dierck.hillmann@uni-bielefeld.de} 


\begin{abstract*} 
Full-field Fourier-domain optical coherence tomography (FF-FD-OCT) enables rapid volumetric imaging but lacks the confocal gating inherent to point-scanning OCT. By acquiring volumes under multiple oblique illuminations, a digital confocal volume can be synthesized in post-processing, a technique originally developed within the framework of reflection matrix imaging (RMI). Existing implementations, however, are often time-domain, off-axis, or use a Michelson interferometer, which reduces sensitivity, speed and the usable space-bandwidth product (SBP). Here, we present an oblique-illumination FF-FD-OCT system in a Mach-Zehnder configuration. This setup enables optimal use of SBP, providing up to a fourfold increase in effective acquisition throughput. In addition, we pose the scattering and aberrations as an overdetermined system, allowing a closed-form solution of an alternating least squares problem. The resulting iterative optimization is equivalent to the state-of-the-art closed-loop accumulation of single scattering (CLASS) algorithm. We demonstrate the setup and reconstruction method in lens tissue, a PSF phantom, a USAF target, and a model eye. The resulting digital confocal volumes provide locally optimized focusing and detection quality, and show improved resolution and signal-to-noise ratio (SNR) compared to conventional FF-FD-OCT. The technique combines the advantages of FF-FD-OCT and confocal OCT while mitigating the limitations of both approaches.
\end{abstract*}

\section{Introduction}
\pagestyle{plain}
Optical coherence tomography (OCT) is a non-invasive, label-free imaging technique used to create volumetric cross-sections of semi-transparent samples with micrometer resolution. Initially conceived in a confocal time-domain (TD) configuration, it used point-scanning and low-coherence interferometry to enable imaging over depth \cite{huang_optical_1991}. Acquisition speed and SNR saw major improvements with the development of Fourier-domain (FD) OCT, which acquired scattering information from all depths simultaneously \cite{boer_improved_2003, leitgeb_performance_2003, yun_high-speed_2003}. A further speed increase was enabled by full-field (FF) OCT, which replaced point-scanning with a global illumination of the area of interest (at the cost of confocality), captured in parallel using a camera  \cite{beaurepaire_full-field_1998, povazay_full-field_2006}. When combined with FD-OCT, the absence of scanning and advancements in camera technology allowed for unprecedented acquisition speeds, making it particularly suitable for certain \textit{in vivo} applications \cite{hillmann_aberration-free_2016}.

Despite these advances in technology, aberrations still pose a challenge in OCT imaging. Aberrations may be induced by both the optical setup and the object (especially deep inside the medium), causing the in- and outgoing waves to be deflected from their ideal single-scattered path \cite{ntziachristos_going_2010, bertolotti_imaging_2022}. The multiply-scattered light thus prevents optimal delivery and collection of light inside the object. Consequently, image quality suffers due to a disturbed point-spread function (PSF) and lower signal-to-noise ratio (SNR). Confocal systems are more resistant to PSF disturbances than full-field systems, since the confocal gate allows partial rejection of out-of-focus and multiply-scattered light \cite{ntziachristos_going_2010, badon_multiple_2017}. However, the confocal gate becomes less effective in the face of strong scattering due to an inability to maintain ideal overlapping illumination and detection pinholes.

To optimize light delivery and detection quality, the field of adaptive optics (AO) has been developed. Here, wavefront shaping is physically implemented using spatial light modulators (SLM) or deformable mirrors in the illumination and/or detection arms \cite{bertolotti_imaging_2022}. It aims to reshape the wavefront to cancel present aberrations. However, this requires the presence of a known guide star (a term originating from astronomy) whose intensity can be optimized, which is often not the case, especially in \textit{in vivo} applications \cite{booth_adaptive_2014, ji_adaptive_2017}. In addition, hardware-based AO generally requires elaborate setups, closed-loop control, and is typically limited to a single isoplanatic patch \cite{park_perspective_2018}.

Computational adaptive optics (CAO) tries to overcome these limitations by correcting for wavefront errors in post-processing. Importantly, this requires the acquired data to be laterally phase-stable. Full-field systems uniquely satisfy this requirement since all lateral points are acquired simultaneously, enabling corrections at every depth and over multiple isoplanatic patches \cite{kumar_numerical_2014, hillmann_aberration-free_2016}. Though CAO is possible in confocal OCT systems \cite{ralston_interferometric_2007, shemonski_computational_2015, adie_computational_2012}, they are inherently sensitive to motion-induced phase errors and may require sophisticated motion-correcting schemes. In its basic implementation, CAO is performed by using a single image and can provide (\textit{in vivo}) aberration-free imaging by optimizing a guide star \cite{adie_guide-star-based_2012, liu_phase_2013}, image metrics \cite{hillmann_aberration-free_2016, thurman_phase-error_2008, adie_computational_2012} or by using split aperture methods \cite{ginner_noniterative_2017, kumar_subaperture_2013, hillmann_computational_2019}.

The most capable form of CAO has been demonstrated using the framework of reflection (and analogous transmission) matrix imaging (RMI), which has been developed over the last decade following work in ultrasound \cite{aubry_detection_2009, lambert_reflection_2020, popoff_controlling_2011, badon_smart_2016}. It stores a set of (complex) wide-field images acquired under different illuminations inside a reflection matrix, which links the inputs and outputs of the imaging system. Essentially, it attempts to acquire the full response from the sample for any input state. Common illumination schemes include point-illumination, (pseudo-)random patterns or, as in this work, oblique illumination (i.e., angled plane waves). Various algorithms can then be used to solve and correct for the scattering behavior, achieving an ideal undisturbed PSF in post-processing and diffraction-limited reconstruction quality.

Most work in RMI has been performed in off-axis holographic setups using low-coherence interferometry, constituting a type of time-domain OCT or microscopy (OCM) \cite{kang_high-resolution_2017, yoon_laser_2020, najar_harnessing_2024}. More recent studies have demonstrated RMI in on-axis swept-source FF-FD-OCT, achieving much higher acquisition speeds than time-domain systems \cite{balondrade_multi-spectral_2024, giraudat_label-free_2026}. Like much preceding work in time-domain, their Michelson interferometer synchronously scans the reference pupil with the illumination pupil. Consequently, the reference wave is tilted on the camera, together with the illumination incident on the object. In time-domain systems, this improves overlap between the object and reference coherence gate, maintaining interference contrast \cite{kim_label-free_2019}. In swept-source FD-OCT, the longer instantaneous temporal coherence length makes the synchronous scanning not strictly necessary and is thus only the result of the choice of setup.

However, a synchronous scan of the reference pupil comes at the cost of space-bandwidth product (SBP). This dimensionless quantity is a measure of the information contained in a measurement, and is the product of the acquired FOV (space) and spatial frequencies (bandwidth) of an object \cite{lohmann_spacebandwidth_1996, park_review_2021}. The tilted reference wave induces a phase gradient on the camera, which increases the detected signal's frequencies. This increase provides no extra information about the object, but does raise the required bandwidth for Nyquist-sampling of the signal. Consequently, not all available bandwidth is used to detect the object's spatial frequencies, resulting in a suboptimal SBP. If bandwidth were used more efficiently, more object information could be retrieved using the same number of camera pixels.

In this paper, we present an oblique illumination FF-FD-OCT system using a fixed on-axis reference wave in a Mach-Zehnder interferometer. This allows the sampling bandwidth to equal the bandwidth of the object signal. By making full use of the available bandwidth, a roughly fourfold improvement in SBP is achieved compared to synchronously scanned-reference pupil setups. Combined with an ultra-fast camera, this leads to an acquisition rate of $18.3$ gigavoxels per second, making it most suitable for \textit{in vivo} applications. We show differences in data-acquisition and processing to a synchronously scanned reference-pupil system and detail our subsequent aberration correction algorithm. The scattering and aberrations are presented as an overdetermined system, allowing a closed-form solution of a least squares problem. Notably, the resulting algorithm is equivalent to the state-of-the-art closed-loop accumulation of single scattering (CLASS) algorithm \cite{kang_high-resolution_2017}. We provide both a mathematical derivation and an intuitive visual explanation of the algorithm.  We quantify resolution improvement with a USAF target. Its capabilities are further demonstrated in a lens tissue, and over depth in a PSF phantom and model eye.

\section{Methods}
\subsection{Setup and data acquisition}

\begin{figure}[htbp]
    \centering
    \includegraphics[width=0.6\textwidth]{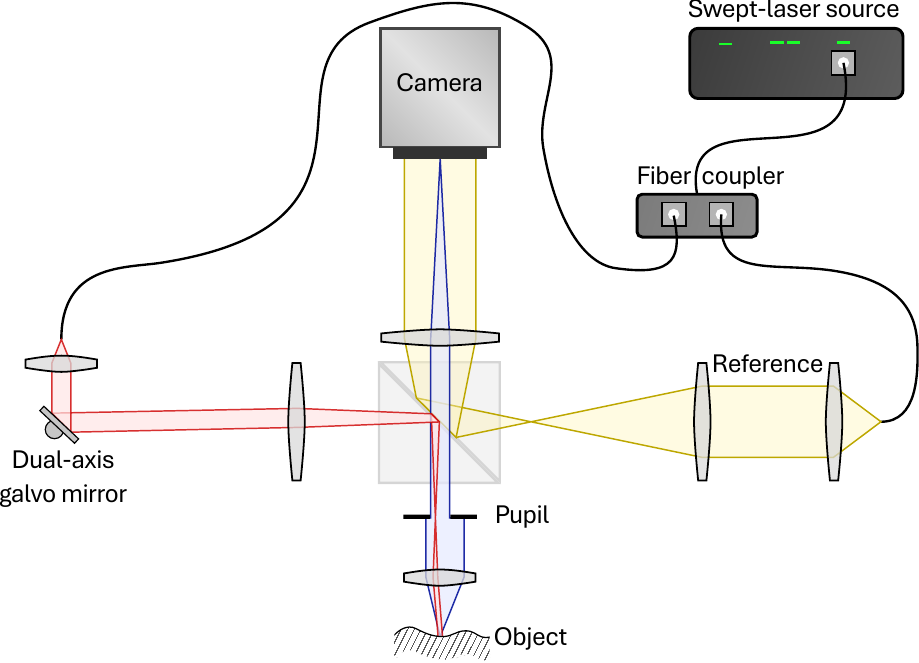}
    \caption{Schematic of the oblique illumination FF-FD-OCT system. Galvo mirrors are imaged onto the sample, allowing the angle of illumination on the object to be changed. The independent reference collimator provides a fixed reference wave on the camera.}
    \label{setup}
\end{figure}

The oblique illumination-capable FF-FD-OCT is configured as a Mach-Zehnder interferometer and is shown in Figure \ref{setup}. Light from the tunable light source (SLDSources.com Broadsweeper BS-840-2-HP) is split by a fiber coupler into object and reference arm. In the object arm, the light is collimated and illuminates a dual-axis galvanometric scanning mirror (ScannerMAX Saturn 9B) that is imaged onto the object using a 4f system. This allows changing the illumination angle on the sample without changing the location of the illuminated area. The backscattered light is imaged via a beam splitter onto the camera. The reference arm contains a separate collimator and the light is recombined via the beam splitter and directed onto the camera. The tunable laser source sweeps from \SI{880}{\nm} to \SI{805}{\nm}. While the laser sweeps, a high-speed camera (Phantom TMX 6410) records the interference between the reference and scattered light from the object. Per sweep and scanner position, $500$ interferograms are recorded using $768 \times 768$ pixels of the sensor at $\numprint{65000}$ fps, at an effective A-line rate of \SI{73}{MHz} or $18.3$ gigavoxels per second. In total $128$ volumes per sample are acquired, each with a different scanner position, using a Fermat spiral pattern \cite{huang_optimization_2014}. In addition, $128$ volumes were acquired for each object without oblique scanning to compare results to conventional FF-FD-OCT. All data were acquired at an NA of $0.13$. The setup is synchronized using an Arduino Giga R1 WiFi, delivering frame-triggers to the camera, triggers to the laser to initiate sweeps, and digital position commands to the scanner between volumes.

\subsection{Detected signal}
In this paper, we use the following conventions: $\vec x$ for the position-space vector $\vec x = (x,y,z)$, $\vec k$ for its Fourier conjugate and $\hat{\vec k}$ for the unit vector corresponding to $\vec k$. Furthermore, $\vec x_\perp$ and $\vec k_\perp$ denote the corresponding projections onto the $x,y$ and $k_x,k_y$ plane, respectively. We furthermore use $\tilde f$ as the Fourier transform of a function $F$.

\subsubsection{OCT reconstruction}
Oblique illumination induces a phase gradient $e^{i \vect{k}^{(i)} \cdot\vect{x}}$ at every depth in the object, where $\vect{k}^{(i)}=k \vect{\hat{k}}^{(i)} \equiv k (\sin{\theta_{x_i}}, \sin{\theta_{y_i}}, \hat k_z^{(i)})$ denotes the $i^{\text{th}}$ illumination vector at illumination angle $\theta_{x_i},\theta_{y_i}$ and wavenumber $k=\frac{2\pi}{\lambda}$. The field detected at the camera plane coming from the object is given by
\begin{align}
    O(\vect{x}_\perp, \vect{k}^{(i)}, k) 
    = P_{\text{in}}(\vect{k}^{(i)}_\perp) \tilde{p}_{\text{out}}(\vect{x_\perp}) *_{\vec x_\perp} A(k) A_O \int \eta(\vect{x'}) \
    \underbrace{\e^{\i \vect{k}^{(i)} \cdot \vect{x'}}}_{\mathclap{e^{i \vect{k}^{(i)}_\perp \cdot \vect{x}_\perp} \e^{\i k^{(i)}_z z'}}}
    \, \e^{\i k_z z'} \, \mathrm{d}z' \text .
    \label{eq:object_arm}
\end{align}
Here, $A(k)$ is the spectrum of the source, $A_O$ the amplitude of the object wave field, $\eta(\vect{x})$ the three-dimensional scattering potential of the object, and $*_{\vec x_\perp}$ denotes the two-dimensional convolution with respect to $\vec x_\perp$. We further define $k^{(i)}_z = \sqrt{k^2-|\vect{k}^{(i)}_\perp|^2}$, $k_z = \sqrt{k^2-|\vect{k}_\perp|^2}$, $\vect{x'}=(x,y,z')=(\vec x_\perp, z')$ and the pupil functions are given by 
\begin{align}
    P_{\text{in/out}}(\vect{k}_\perp)=
    \begin{cases} 
        \e^{\i \phi_{\text{in/out}}(\vect{k}_\perp)} & \text{for $|\vec k_\perp| < k \mathrm{NA}_{\text{in/out}}$ } \\
        0 & \text{otherwise, } 
    \end{cases}
\end{align}
with $\phi_\text{in/out}$ describing the phase error of the input and output aberration, and $\mathrm{NA}_\text{in/out}$ describing the numerical aperture (NA) of the input and output. While $P_\text{in}$ and $P_\text{out}$ should in principle be identical, misalignment and motion-induced global phase changes can show up as input aberrations, making a separation crucial. The reference wave is delivered by a fixed reference collimator with amplitude $A_R$, at a total distance $z_0$ corresponding to the zero-delay line:
\begin{align}
    R(k) &= A(k) A_R \; \e^{\i k z_0} \text .
\end{align}

In FD-OCT, interferograms $I$ are recorded as a function of the wavenumber $k$. This gives the expression 
\begin{align}
    I(\vect{x}_\perp, \vect{k}^{(i)}, k) &= |O(\vect{x}_\perp, \vect{k}^{(i)}, k) + R(k)|^2 \\
    &= |O|^2(\vect{x}_\perp, \vect{k}^{(i)}, k) + |R|^2(k) + (O R^*)(\vect{x}_\perp, \vect{k}^{(i)}, k)  + (RO^*)(\vect{x}_\perp, \vect{k}^{(i)}, k) \text.
\end{align}
The first two terms are the autocorrelation and DC terms. The last two terms are the signal and the conjugate signal. Concentrating on the signal term and assuming, without loss of generality that $z_0=0$, this becomes
\begin{align}
    O R^*(\vect{x}_\perp, \vect{k}^{(i)}, k) = P_{\text{in}}(\vect{k}^{(i)}_\perp) \tilde p_\text{out}(\vec x_\perp) *_{\vec x_\perp} S(k) \int \eta(\vec x_\perp, z') \e^{\i \vec k_\perp^{(i)} \cdot \vec x_\perp} \e^{\i (k_z + k_z^{(i)}) z'} \d z' \text, \label{eq:oct_signal}
\end{align}
with $S(k) = A_R A_O |A|^2(k)$.

Taking the lateral Fourier transform of the signal term $O R^*$ and applying the convolution theorem as well as the Fourier shift theorem reveals that the object's lateral $\vect{k}_\perp$-space shifts inside the detection pupil as the illumination is rotated:
\begin{align}
    \mathcal{F}_{\vect{x}_\perp\rightarrow \vect{k}_\perp}\{O R^* (\vect{x}_\perp, \vect{k}^{(i)}_\perp, k)\} 
    &= P_{\text{in}}(\vect{k}^{(i)}_\perp) P_{\text{out}}(\vect{k_\perp}) S(k)  \int \tilde{\eta}(\vect{k}_\perp-\vect{k}^{(i)}_\perp, z') \e^{\i (k_z + k_z^{(i)}) z'} \mathrm{d}z' \text. \label{eq:lateral_shift}
\end{align}
With $P_\text{out}$ describing a constant band-pass and $\tilde \eta(\vec{k}_\perp - \vec{k}^{(i)}_\perp, z)$ being shifted by the illumination $\vec{k}^{(i)}_\perp$, the setup acquires spatial frequencies that are inaccessible with conventional full-field setups \cite{heintzmann_super-resolution_2017}.
OCT depth-reconstruction is performed by Fourier transforming $O R^*$ along the $k$-axis. Assuming the NA is sufficiently small such that $k_z \approx k$ and $k^{(i)}_z \approx k$, we obtain
\begin{align}
    \mathcal{F}_{k \rightarrow z}\{O R^*(\vect{x}_\perp, \vect{k}^{(i)}_\perp, k)\} 
    &\approx P_{\text{in}}(\vect{k}^{(i)}_\perp) \tilde p_\text{out}(\vec x_\perp) *_{\vec x_\perp} S(k) \iint \eta(\vec x_\perp, z') e^{ik (\vect{\hat{k}}^{(i)}_\perp \cdot \vect{x}_\perp + 2 z' - z) }\mathrm{d}k\ \mathrm{d}z'\ \\
    &= P_{\text{in}}(\vect{k}^{(i)}_\perp) \tilde p_\text{out}(\vec x_\perp) *_{\vec x_\perp} \tilde{s}(z) *_z \eta\left(\vect{x}_\perp, \frac{1}{2}(z-\vect{\hat{k}}^{(i)}_\perp \cdot \vect{x}_\perp)\right) \text, \label{eq:oct_rec}
\end{align}
where $\tilde{s}(z)\equiv \mathcal{F}\{S(k)\}$ is the axial PSF. Eq.~\eqref{eq:oct_rec} gives a sheared version of the object's scattering potential, i.e., different lateral locations from the same layer of the object will be shifted axially by $\hat{\vec{k}}^{i}_\perp \cdot \vec x_\perp$. Physically, this arises due to lateral differences in optical path length (OPL) between the object and reference wave at the camera plane, induced by the oblique illumination. 

To compensate for this effect, we applied the conjugate of the original phase gradient $\exp(-\i \vec{k}^{(i)}_\perp \cdot \vect{x}_\perp )$ to the axial spectrum Eq.~\eqref{eq:oct_signal} before the Fourier transform along the $k$-axis:
\begin{align}
    U(\vect{x}, \vect{k}^{(i)}_\perp) &= \mathcal{F}_{k\rightarrow z}\left\{ O R^*(\vect{x}_\perp, \vect{k}^{(i)}_\perp, k) \cdot \e^{-\i k \vect{\hat{k}}^{(i)}_\perp \cdot \vect{x}_\perp } \right\}\\
    &\approx P_{\text{in}}(\vect{k}^{(i)}_\perp) \tilde p_\text{out}(\vec x_\perp)e^{-i \vect{k}^{(i)}_\perp \cdot \vect{x}_\perp } *_{\vec x_\perp}  \tilde{s}(z) *_z \eta\left(\vect{x}_\perp, \frac{z}{2}\right)
    \label{eq:shear_correction} \text .
\end{align}
This step also compensates the lateral shift in the object's $\vec k_\perp$-space (Eq.~\eqref{eq:lateral_shift}), aligning it between measurements.

By summing over the shear-corrected fields $U$ in post-processing, we create what we will refer to as a digital confocal volume,
\begin{align}
    U_{\text{Dig.Conf.}}(\vect{x}) &= \sum_{i} U(\vect{x}, \vect{k}^{(i)}_\perp) \\
    &\approx \tilde p_{\text{in}}(-\vec x_\perp) \tilde p_\text{out}(\vec x_\perp) *_{\vec x_\perp}  \tilde{s}(z) *_z \eta\left(\vect{x}_\perp, \frac{z}{2}\right) \text, \label{eq:digconf}
\end{align}
where we used the approximation $\sum_i P_{\text{in}}(\vect{k}^{(i)}_\perp) e^{-i \vect{\hat{k}}^{(i)}_\perp \cdot \vect{x}_\perp } \approx \tilde p_{\text{in}}(-\vec x_\perp)$ that would become accurate if $\vec k_\perp^{(i)}$ densely scanned the pupil. The summation over plane wave illuminations creates a virtual illumination spot at every point of the resulting volume. At the same time, the camera pixels approximate point-detection. These two combine to create a confocal filter in post-processing. The resulting PSF is now the multiplication of the (virtual) mirrored input and output PSF, improving image quality accordingly. In practice, the limited number of illuminations prevents focusing to a single spot. More importantly however, aberrations need to be addressed to obtain optimal overlapping virtual illumination and detection pinholes and achieve maximal image quality.

\subsection{Aberration correction}
A single acquired field obtained for a single incident wave vector $\vec k^{(i)}_\perp$ contains a mix of both object information $\eta$ and aberrations $P_\text{in/out}$. This makes the problem of separating the two from a single field fundamentally underdetermined and ill-posed. In our full data, each illumination provides partially overlapping object information while aberrations remain constant, effectively overdetermining the system. In principle, this allows us to separate and solve for object and aberrations objectively, requiring no assumptions or priors. In the following, we omit the depth-index $z$ for readability. We can formulate the solution as a least squares problem in $\vect{k}_\perp$-space:

\begin{align}
    \hat P_\text{in}, \hat P_\text{out}, \hat{\tilde \eta} = \argmin_{P_\text{in}, P_\text{out}, \tilde \eta} \sum_i \sum_{\vect{k}_\perp} \left| \tilde{u}_i^{\text{obs}}(\vect{k}_\perp) - A \tilde{\eta}(\vect{k}_\perp) P_{\text{in}}(\vect{k}^{(i)}_\perp) P_{\text{out}}(\vect{k}_\perp+\vect{k}^{(i)}_\perp) \right|^2,
    \label{loss_func}
\end{align}
Here, $\tilde u_i^{\text{obs}}$ denotes the lateral Fourier transform of the measured fields (from hereon referred to as apertures) after shear correction, functioning as the observations for the model. These apertures are described in terms of a constant amplitude $A$, and the unknown scattering potential $\eta$, and pupils $P_\text{in}$ and $P_\text{out}$ to be determined via optimization. The model assumes single scattering.

An efficient way to determine the unknown functions is to use alternating least squares: setting the gradients with respect to $\tilde \eta(\vec k_\perp)$, $P_\text{in}(\vec k_\perp^{(i)})$, $P_\text{out}(\vec k_\perp)$ to $0$ while keeping the others fixed and solving for the respective parameters one after the other. Since this analytically determines the extremum in a subspace, it will iterate towards the optimum. This leads to the following update rules, where $j$ is the current iteration:

\begin{align}
    \tilde{\eta}_{j+1}(\vect{k}_{\perp}) &\leftarrow 
    \dfrac{\sum\limits_i \tilde{u}_i^{\text{obs}}(\vect{k}_{\perp}) \, P_{\text{in}, j}^*(\vect{k}^{(i)}_{\perp}) \, P_{\text{out}, j}^*(\vect{k}_{\perp} + \vect{k}^{(i)}_{\perp})}
    {A \sum\limits_i \left| P_{\text{in}, j}(\vect{k}^{(i)}_{\perp}) \right|^2 \, \left| P_{\text{out}, j}(\vect{k}_{\perp} + \vect{k}^{(i)}_{\perp}) \right|^2} 
    \label{eq:est_eta} \\[15pt]
P_{\text{in}, j+1}(\vect{k}^{(i)}_{\perp}) &\leftarrow 
    \dfrac{\sum\limits_{\vect{k}_{\perp}} \tilde{u}_i^{\text{obs}}(\vect{k}_{\perp} - \vect{k}^{(i)}_{\perp}) \, P_{\text{out}, j}^*(\vect{k}_{\perp}) \, \tilde{\eta}_{j+1}^*(\vect{k}_{\perp} - \vect{k}^{(i)}_{\perp})}
    {A \sum\limits_{\vect{k}_{\perp}} \left| \tilde{\eta}_{j+1}(\vect{k}_{\perp} - \vect{k}^{(i)}_{\perp}) \right|^2 \, \left| P_{\text{out}, j}(\vect{k}_{\perp}) \right|^2} 
    \label{eq:est_pin} \\[15pt]
P_{\text{out}, j+1}(\vect{k}_{\perp}) &\leftarrow 
    \dfrac{\sum\limits_i \tilde{u}_i^{\text{obs}}(\vect{k}_{\perp} - \vect{k}^{(i)}_{\perp}) \, P_{\text{in}, j+1}^*(\vect{k}^{(i)}_{\perp}) \, \tilde{\eta}_{j+1}^*(\vect{k}_{\perp} - \vect{k}^{(i)}_{\perp})}
    {A \sum\limits_i \left| \tilde{\eta}_{j+1}(\vect{k}_{\perp} - \vect{k}^{(i)}_{\perp}) \right|^2 \, \left| P_{\text{in}, j+1}(\vect{k}^{(i)}_{\perp}) \right|^2} 
    \label{eq:est_pout}
\end{align}

In practice, it was found that modifying Eq.~\ref{eq:est_pin} such that the output pupil correction was not applied led to better convergence for scattering samples. This is likely due to the large number of degrees of freedom in $P_{\text{out}}$ strongly affecting the small number of degrees of freedom in $P_{\text{in}}$, which may be problematic if initial estimates are noisy. The following form was therefore used in practice for scattering samples:
\begin{align}
   P_{\text{in}, j+1}(\vect{k}^{(i)}_{\perp}) &\leftarrow 
    \dfrac{\sum\limits_{\vect{k}_{\perp}} \tilde{u}_i^{\text{obs}}(\vect{k}_{\perp} - \vect{k}^{(i)}_{\perp}) \, \tilde{\eta}_{j+1}^*(\vect{k}_{\perp} - \vect{k}^{(i)}_{\perp})}
    {A \sum\limits_{\vect{k}_{\perp}} \left| \tilde{\eta}_{j+1}(\vect{k}_{\perp} - \vect{k}^{(i)}_{\perp}) \right|^2 \, \left| P_{\text{out}, j}(\vect{k}_{\perp}) \right|^2} \text. 
\end{align}
In addition, we omit the normalization in the denominators. This leads to a weighting of $\tilde{\eta}_{j+1}$ based on the number of apertures covering each pixel. This makes the subsequent determination of the pupil functions less sensitive to low-amplitude phasors. In addition, it leads to an apodization which suppresses higher frequencies and approximates the transfer function of a confocal system. For the pupil functions, we ultimately correct the observed apertures using the phase functions $\phi_{\text{in/out,j+1}}(\vect{k}_\perp) = \arg P_{\text{in/out}, j+1}(\vect{k}_\perp)$, whose phase is unaffected by normalization of the pupil functions. 

\begin{figure}[htpb]
    \centering
    \includegraphics[width=1\linewidth]{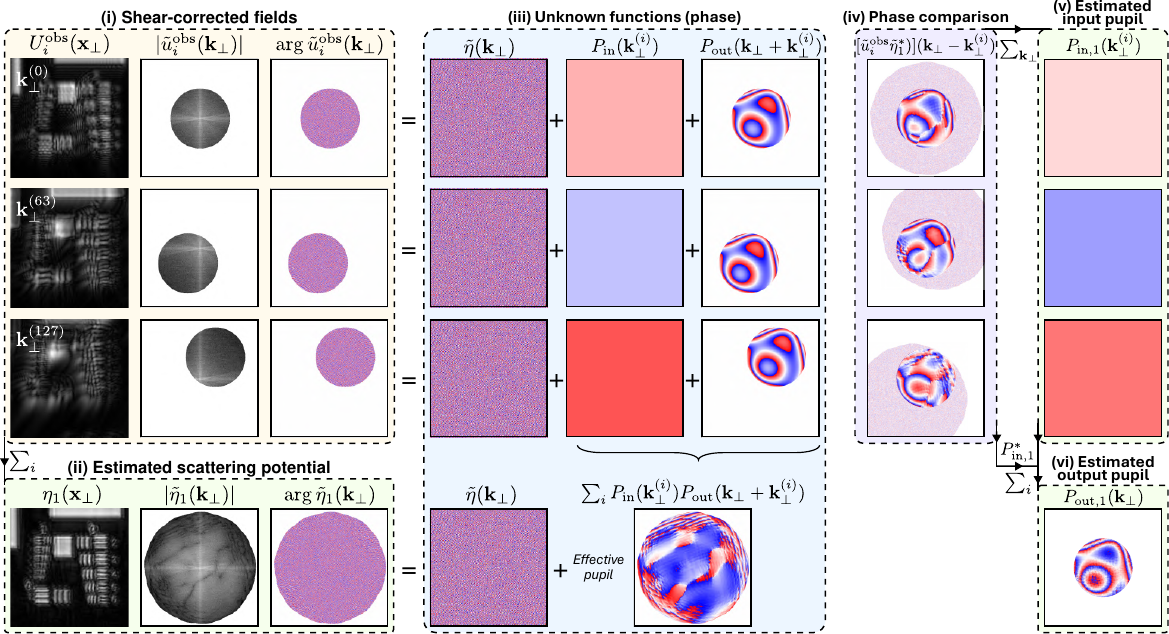}
    \caption{Schematic showing the principle behind estimating object $\eta$ and aberrations during the first iteration. Shear-corrected apertures $\tilde u^{\text{obs}}_i(\vec k_\perp)$ (i)  are summed to create the estimate $\tilde{\eta}_1$ (ii). Its effective pupil shows less aberrated phase than the individual pupils (iii). Subsequent phase-comparison between the apertures and $\tilde{\eta}_1$ shows the difference in aberrations (iv). Summing over $\vec k_\perp$ estimates $P_{\text{in},1}$ (v), while summing over the input-corrected illuminations $i$ estimates $P_{\text{out},1}$ (vi).}
    \label{fig::ab_cor}
\end{figure}

Figure \ref{fig::ab_cor} shows a schematic representation of how the estimates $\tilde{\eta}_1$, $P_{\text{in},1}$ and $P_{\text{out},1}$ are determined during the first iteration when no initial aberration correction is applied. We start at (\textit{i}) with the shear-corrected fields. Per illustration, one on-axis ($\vect{k}^{(0)}_\perp=\vect{0}$) and two oblique images are shown. Their corresponding apertures are aligned in zero-padded $\vect{k}_\perp$-space. The best estimate of the scattering potential $\tilde{\eta}_1$ (Eq.~\eqref{eq:est_eta} for $j=0$) is created in (\textit{ii}) by summing over all the illuminations. Notably, $\tilde{\eta}_1$ is the lateral Fourier transform of the digital confocal image in Eq.~\eqref{eq:digconf}, showing a larger NA than the individual apertures. The corresponding image $\eta_1$ appears less aberrated than any individual image $U_i^{\text{obs}}$ due to the multiplication of the output and input PSF suppressing sidelobes. In (\textit{iii}) the underlying unknown functions are shown which make up our observed fields and which we try to estimate. The underlying effective pupil function of $\tilde{\eta}_1$ is the sum of the shifted pupil functions: $\sum_i P_{\text{out}}(\vect{k}_\perp+\vect{k}^{(i)}_\perp)P_{\text{in}}(\vect{k}^{(i)}_\perp)$. This summation averages out fluctuations in the aberrated phase of the individual apertures, equivalently explaining why $\tilde{\eta}_1$ appears less aberrated.

To estimate the aberrations, we first compare the phases of the apertures to $\tilde{\eta}_1$ in (\textit{iv}). Since the underlying object's $\vect{k}_\perp$-space $\tilde{\eta}$ is identical (as seen in (\textit{iii})), we observe only the phase difference between the aberrations of the apertures and the effective pupil of $\tilde{\eta}_1$. Critically, since the aberrations in the effective pupil appear roughly constant in phase, the differences approximate the output aberration, up to global phase offsets determined by the input aberrations. In (\textit{v}), we can therefore estimate $P_{1,\text{in}}$ by summing over phasors $\vect{k}_\perp$ (Eq.~\eqref{eq:est_pin}). We then estimate $P_{1,\text{out}}$ in (\textit{vi}) by summing over the illuminations $i$ instead, using $P^*_{1,\text{in}}$ to remove the global phase offsets (Eq.~\eqref{eq:est_pout}). Note that for this summation, the phase differences must be centered.

By correcting the apertures using these estimates during the next iteration, the effective aberration of $\tilde{\eta}_{2}$ is further reduced, improving the estimates of $P_{\text{in/out},2}$ in turn. The process is repeated until convergence is reached. The optimization can be performed for smaller patches in case of anisoplanatic scattering. The phase of $P_{\text{in/out}}$ can only be determined up to a piston, tip and tilt\cite{kang_high-resolution_2017, lambert_ultrasound_2022}, since any gradient in $P_\text{out}$ can be compensated by a gradient in $P_\text{in}$, resulting in the same loss value. To prevent stitching artifacts between patches, these gradients are removed by centering the center-of-mass of $|\tilde{p}_{j,\text{out}}|^4$.

\subsubsection{$\vec k_\perp^{(i)}$-registration}
The values of $\vec k_\perp^{(i)}$ need to be known accurately to allow for optimal convergence and image formation. We approach this as a registration problem, treating the apertures $\tilde{u}^{\text{obs}}_i$ at a common depth as laterally shifted images. This avoids relying on calibration measurements, which may not be possible in \textit{in vivo} imaging.
First, an initial shear compensation is applied using theoretical values derived from system specifications and scanner lookup table. Volumes are then axially registered to correct path-length variations induced by dual-axis galvo scanning, which lacks a single pivot point. Apertures are subsequently extracted from a common layer showing strong signal, typically near the surface. Relative shifts for all apertures are then determined using phase correlation on the amplitudes only. The shift values are optimized using a weighted least squares fit, similar to GraphSLAM \cite{thrun_graph_2006}. Weights are given by the phase correlation peak value, or set to zero if deviations exceed $15$ pixels of the relative theoretical shifts.

\subsection{Comparison to synchronously scanned reference pupil}
Previous work in oblique illumination using on-axis FF-FD-OCT placed a mirror in the reference arm, creating a Michelson interferometer. This causes the reference pupil to be scanned synchronously with the object pupil \cite{balondrade_multi-spectral_2024}, such that the reference wave induces a phase gradient on the camera (Fig.~\ref{fig:fix_vs_sync}b).

\subsubsection{Space-bandwidth product}
\begin{figure}[htpb]
\centering
\includegraphics[width=1\textwidth]{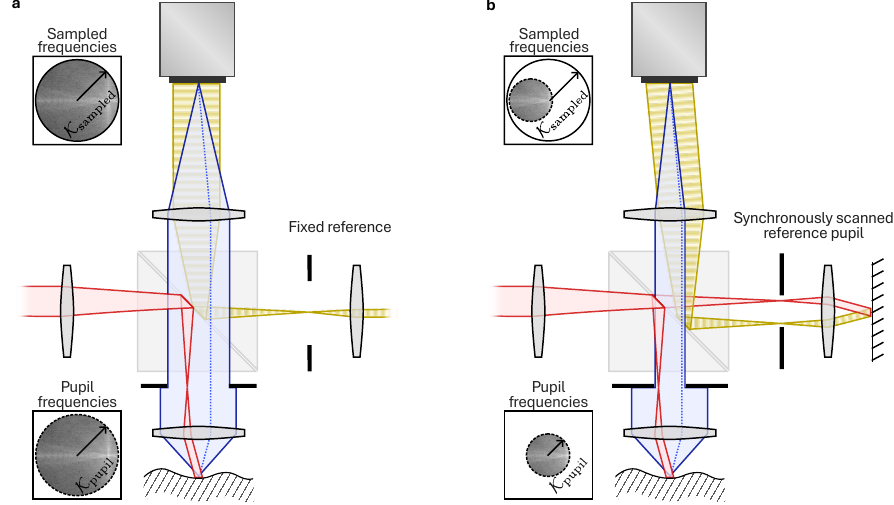}
\caption{Schematic comparing a system with a fixed reference arm (ours, \textbf{a}) to a synchronously scanned reference pupil (\cite{balondrade_multi-spectral_2024}, \textbf{b}). In the latter, the phase gradient from the reference on the camera shifts the spatial frequencies after passing through the detection pupil. This increases the required sampling bandwidth, requiring the detection NA to be lowered to prevent aliasing. DC-component of scattered field shown in blue dotted line.}
\label{fig:fix_vs_sync}
\end{figure}

This Michelson configuration leads to a much lower SBP compared to a fixed reference. Fig.~\ref{fig:fix_vs_sync} compares the setup and sampled frequencies of a setup using a fixed reference and a synchronously scanned reference. The sampling bandwidth is defined by $\mathcal{K}_{\text{sampled}} = \frac{\pi}{\Delta x} M$, with $\Delta x$ the pixel size of the camera and $M$ the magnification from object to camera. The detection pupil transmits frequencies up to $\mathcal{K_{\text{pupil}}} = k \text{NA}_{\text{det}}$. A fixed on-axis reference does not modulate the detected frequencies, requiring $\mathcal{K}_{\text{sampled,fix}} \geq \mathcal{K_{\text{pupil}}}$ for sampling at Nyquist (Fig.~\ref{fig:fix_vs_sync}a). This allows optimal use of the available bandwidth. 

In contrast, the added phase gradient in a synchronously scanned pupil adds a carrier frequency to the frequencies passing through the pupil (Fig.~\ref{fig:fix_vs_sync}b). This increases the required sampling bandwidth to $\mathcal{K}_{\text{sampled,sync}} \geq k \left( \text{NA}_{\text{illum}}+\text{NA}_{\text{detect}} \right)$, with a full scan of the pupil requiring $\mathcal{K}_{\text{sampled,sync}} \geq 2\mathcal{K_{\text{pupil}}}$ in both axes. This requirement can be fulfilled by either decreasing $\text{NA}_{\text{det/illum}}$ or FOV by a factor of two in each direction. In either case, a synchronously scanned reference pupil reduces SBP by a factor of four compared to our setup, requiring four times more pixels in each camera image compared to our methods. In practice, this will reduce imaging speed by a factor of four assuming a constant pixel digitization rate. 

\subsubsection{Reconstruction}
When scanning the reference pupil, the phase gradient on the camera is $\vect{k}^{(i)}_\perp$-dependent, given by:

\begin{align}
    R_{\text{sync}}(\vect{k}^{(i)}_\perp, k) &= A(k) A_R \ e^{\i kz_0}\underbrace{e^{\i \vect{k}^{(i)}_\perp\cdot\vect{x}_\perp}}_{\mathclap{\text{scan-induced phase gradient}}} e^{\i k^{(i)}_z z_0} \text.
\end{align}
At the same time the object arm is the same as in Eq.~\eqref{eq:object_arm}. The object and reference now maintain a common zero-delay line, such that the OCT reconstruction does not appear sheared when transforming over the $k$-axis.

\begin{align}
    &\mathcal{F}_{k\rightarrow z}\{O R^*_{\text{sync}}(\vect{x}_\perp, \vect{k}^{(i)}, k)\} \notag \\ &\quad =  \int S(k) \left[ P_{\text{in}}(\vect{k}^{(i)}_\perp) \tilde p_\text{out}(\vec x_\perp) *_{\vec x_\perp} \e^{\i k \vect{\hat{k}}_\perp^{(i)} \cdot \vec x_\perp} \int \eta(\vec x_\perp, z')  \e^{\i (k_z + k_z^{(i)}) z'} \d z' \right] e^{-\i k(\vect{\hat{k}}^{(i)}_\perp\cdot\vect{x}_\perp - z)} \mathrm{d} k \\
    &\quad\approx P_{\text{in}}(\vect{k}^{(i)}_\perp) \tilde p_\text{out}(\vec x_\perp)e^{-\i \vect{\hat{k}}^{(i)}_\perp \cdot \vect{x}_\perp } *_{\vec x_\perp}  \tilde{s}(z) *_z \eta\left(\vect{x}_\perp, \frac{z}{2}\right) \label{sync_rec} \text.
\end{align}
Here, the integral in square brackets is the object term, which only interacts with the reference field after having passed through the detection pupil. The final reconstruction in Eq.~\eqref{sync_rec} is identical to our shear-corrected reconstruction in Eq.~\eqref{eq:shear_correction}. Our shear correction is thus the computational counterpart to physically scanning the reference pupil. However, whereas the synchronous scanning requires a larger sampling bandwidth compared to the detection pupil to prevent aliasing, we can zero-pad our $\vect{k}_\perp$-spaces in post-processing to allow the apertures to shift without aliasing. We thus arrive at the same aligned apertures, however, containing up to four times the FOV or resolution.

\subsection{Relation to CLASS}
The aberration correction presented here is equivalent to CLASS \cite{kang_high-resolution_2017}. CLASS aims to maximize the coherent sum of the aberration-corrected fields, showing that it is only maximal when all aberrations have been fully corrected \cite{kang_high-resolution_2017}. This method was introduced to improve the previously-developed collective accumulation of single-scattered waves (CASS) image, which summed the fields without performing any corrections \cite{kang_imaging_2015}. Our initial unoptimized estimate $\eta_1$ (Eq.~\eqref{eq:est_eta} for $j=0$, which has no corrections applied) thus equals the CASS image. We then correct the observed fields $\tilde{u}_i^{\text{obs}}$ such that we minimize the difference with respect to the common object $\eta_1$. This (partially) phase-aligns the corrected observed fields, since if they were not phase-aligned after correction, they would not correspond to the same object. However, the sum of coherent corrected fields is only maximal if they are phase-aligned. Therefore, both methods ultimately arrive at the same optimization procedure, with the CLASS-image being the improved estimate $\tilde{\eta}_{j+1}$ (Eq.~\eqref{eq:est_eta}) \cite{yoon_laser_2020}. 

\section{Results}
\begin{figure}[htpb]
\centering
\includegraphics[width=1\textwidth]{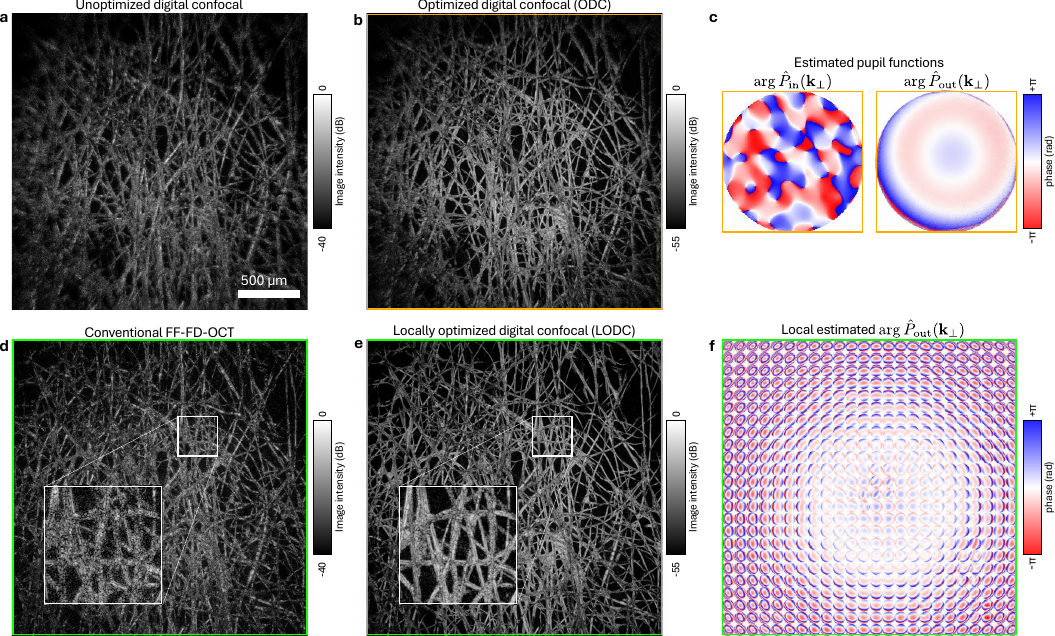}
\caption{Lens tissue. \textbf{a,b.} Digital confocal image before and after optimization. \textbf{c.} Corresponding estimated input and output pupil functions used for \textbf{b}. \textbf{d.} Coherent average of locally corrected on-axis volumes \textbf{e.} Locally optimized digital confocal image. \textbf{f.} Local estimated input and output pupil functions.}
\label{fig:lens_tissue}
\end{figure}

We first demonstrate the least squares optimization in a lens tissue (MC-5, Thorlabs). Fig.~\ref{fig:lens_tissue}a shows the unoptimized digital confocal image $\eta_{1}$. The optimization typically converged after a couple of iterations (at which point aberrations were unchanging), but was run for 100 iterations globally and 20 locally. After iterating, the optimized digital confocal (ODC) enface shows improved image quality and higher SNR (Fig.~\ref{fig:lens_tissue}b, note the change in dynamic range shown in the image). The corresponding determined aberrations are shown in Fig.~\ref{fig:lens_tissue}c. Although $\hat{P}_{\text{out}}$ shows only a slight defocus, $\hat{P}_{\text{in}}$ appears speckle-like. By optimizing over smaller isoplanatic patches ($198 \times 198$~$\mu$m, $49$~$\mu$m overlap) we obtain a locally optimized digital confocal (LODC) enface, showing consistent image quality over the full FOV (Fig.~\ref{fig:lens_tissue}e). The corresponding spatially varying output aberrations are displayed in Fig.~\ref{fig:lens_tissue}f. Compared to a conventional FF-FD-OCT, consisting of a phase-aligned, coherent average of on-axis volumes having undergone the same corrections (Fig.~\ref{fig:lens_tissue}d), the (L)ODC enface shows strong improvement in both resolution and SNR.

\begin{figure}[htpb]
\centering
\includegraphics[width=1\textwidth]{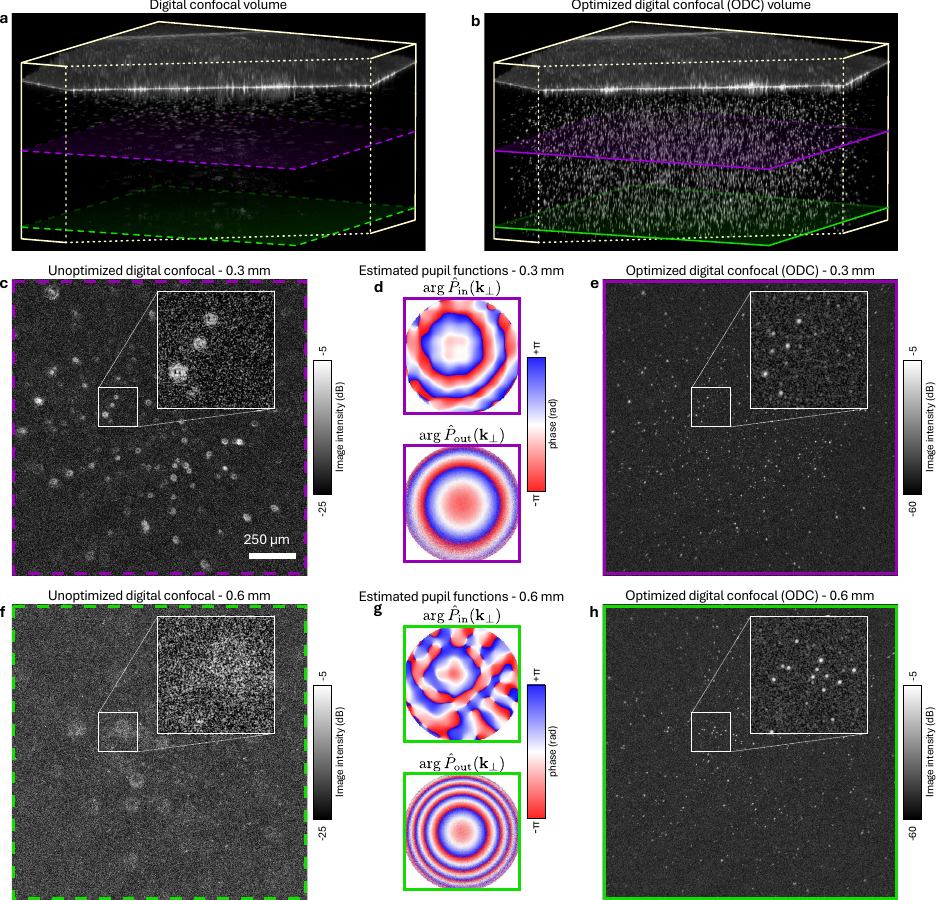}
\caption{PSF phantom. \textbf{a.} 3D-sliced digital confocal volume, optimized using only the top layer. \textbf{b.} 3D-sliced digital confocal volume optimized at every depth. \textbf{c,f} Unoptimized enfaces at depths of \SI{0.3}{\mm} and \SI{0.6}{\mm} of volume shown in \textbf{a.}. \textbf{e,h} ODC enfaces at depths of \SI{0.3}{\mm} and \SI{0.6}{\mm} of volume shown in \textbf{b.}. \textbf{d, g.} Estimated input and output pupil functions at depths \SI{0.3}{\mm} and \SI{0.6}{\mm}, respectively.}
\label{fig:psf_phantom}
\end{figure}

Next, we evaluated performance over depth by imaging a PSF phantom (OCTPSF01, BioPixS). Input and output aberrations were first estimated at the surface, and the corresponding corrections were subsequently applied to every depth of the volume. A sliced 3D view of the resulting ODC volume is shown in Fig.~\ref{fig:psf_phantom}a. Though strong signal is now obtained from the surface, point scatterers blur with increasing depth, as seen in the unoptimized enfaces in Fig.~\ref{fig:psf_phantom}c and f. The volumes were then corrected sequentially from the surface downwards, using the pupil functions estimated at depth $z$ as initial estimates for depth $z+1$. The recovered input and output aberrations at in-medium (refractive index $n=1.5$) depths of \SI{0.3}{\mm} and \SI{0.6}{\mm} are shown in Figs.~\ref{fig:psf_phantom}d and \ref{fig:psf_phantom}g, respectively, showing a depth-dependent defocus. The off-center defocus in the input aberration indicates a misalignment between the illumination and detection pupil. After correction, strong signal is recovered throughout the ODC volume (Fig.~\ref{fig:psf_phantom}b), with scatterers now resolvable at all depths (Figs.~\ref{fig:psf_phantom}e and h).

We investigated potential translation to retina by imaging a model eye (Modell-Augen Manufaktur). For this acquisition, the Fourier transform of the detected object field was set to coincide with the model eye's pupil since most aberrations are expected to occur there. This was accomplished by moving the lens in the reference arm, inducing a quadratic phase on the image plane, and propagating the Fourier transform in turn. Any residual misalignment between the two planes is corrected computationally \cite{kwon_computational_2023}. We determined the computational propagation distance by running the algorithm for several values and subsequently selecting the value that maximizes the total energy of the ODC image $\sum_{x,y} |\hat{\eta}(x,y)|^2$. 

\begin{figure}[htpb]
\centering
\includegraphics[width=1\textwidth]{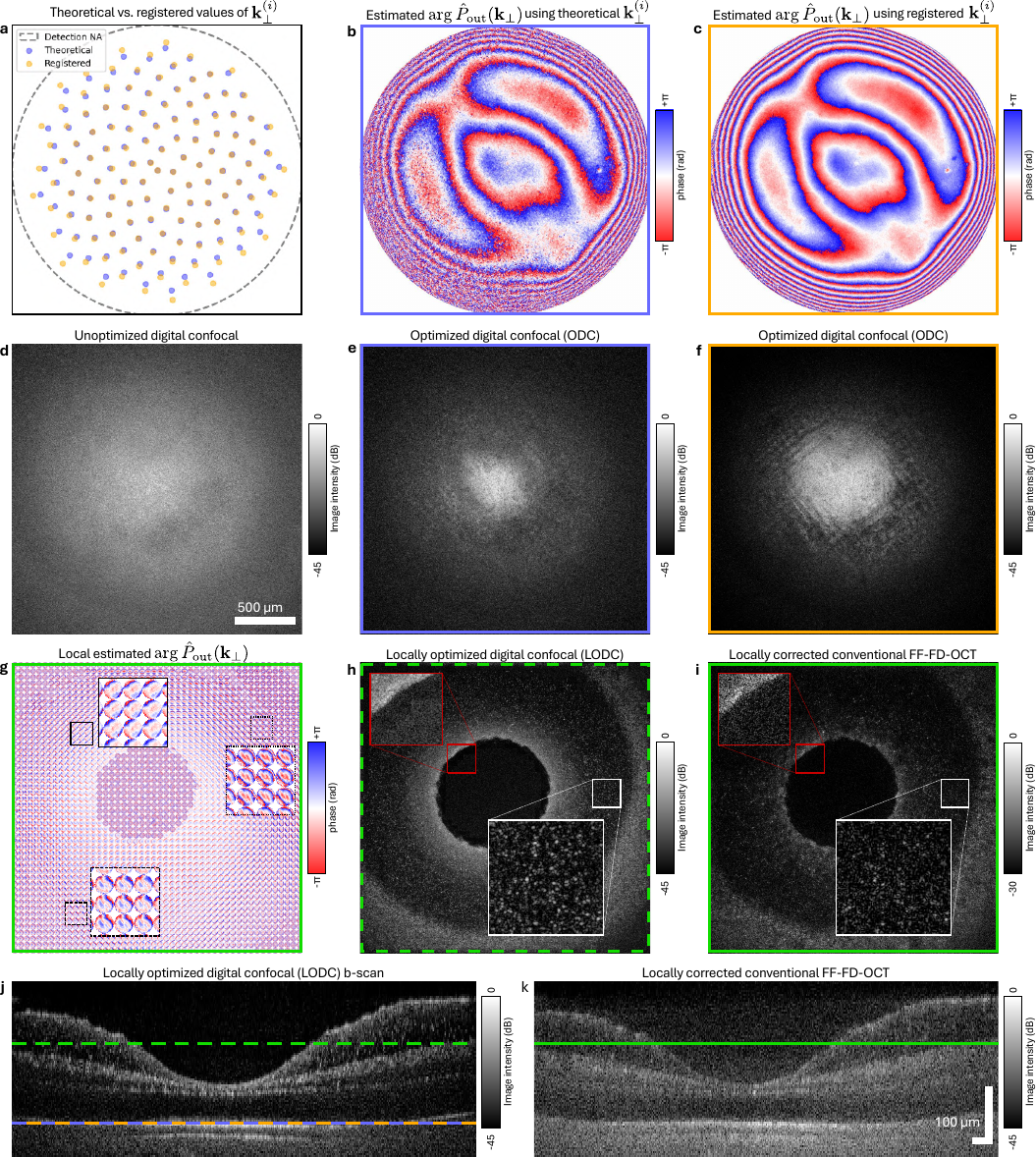}
\caption{Model eye. \textbf{a.} Comparison of theoretical and registered $\vect{k}^{(i)}_\perp$ values. \textbf{b,c.} Output pupil functions estimated using the theoretical and registered $\vect{k}^{(i)}_\perp$ values, respectively, using layer shown in \textbf{d}. \textbf{e,f.} Corresponding optimized digital confocal (ODC) enfaces using theoretical and registered $\vect{k}^{(i)}_\perp$ values. \textbf{g.} Estimated local output pupils for ODC enface shown in \textbf{h}. \textbf{i.} Correction shown in \textbf{g.} applied to conventional FF-FD-OCT enface. \textbf{j,k.} B-scan of locally optimized digital confocal (LODC) volume and same corrections applied to conventional FF-FD-OCT volume.}
\label{fig:model_eye}
\end{figure}

Figure \ref{fig:model_eye}a compares the theoretical values of $\vect{k}^{(i)}_\perp$ based on the scan protocol to the values obtained using our $\vect{k}^{(i)}_\perp$-registration. Figures \ref{fig:model_eye}b and c show the output pupil functions estimated using the most reflective layer shown (marked in Fig.~\ref{fig:model_eye}j), using the theoretical values and registered values of $\vect{k}^{(i)}_\perp$, respectively. The resolution in the estimated aberration is improved when using the registered values. The unoptimized enface is shown in Fig.~\ref{fig:model_eye}d. The ODC enface optimized using the registered $\vect{k}^{(i)}_\perp$ (Fig.~\ref{fig:model_eye}f) shows improved SNR and reveals the model eye's texture compared to using the theoretical values (Fig.~\ref{fig:model_eye}e). The estimated input and output pupil functions were applied to each layer and the volumes subsequently refocused over depth. To this end, we manually determined the quadratic phase required to correct defocus at a distant layer and interpolated this correction to all other layers.

Next, we iterated through all layers and performed local corrections over smaller isoplanatic patches ($99~\text{\textmu m} \times 99~\text{\textmu m}$, $25~\text{\textmu m}$ overlap). Figure \ref{fig:model_eye}g shows one such map of local output aberrations varying over the FOV. Figure \ref{fig:model_eye}h shows the corresponding LODC enface. The local optimization allows us to more clearly resolve scatterers (white inset), but pulls up noise in empty regions (red inset, displayed with increased dynamic range). Figure \ref{fig:model_eye}i shows the same corrections applied to the conventional FF-FD-OCT data. The LODC image has better SNR and reveals scatterers that are otherwise not visible. Similar improvements can be seen when comparing an LODC B-scan (Fig.~\ref{fig:model_eye}j) to a conventional FF-FD-OCT B-scan having undergone the same corrections (Fig.~\ref{fig:model_eye}k).

\begin{figure}[htpb]
\centering
\includegraphics{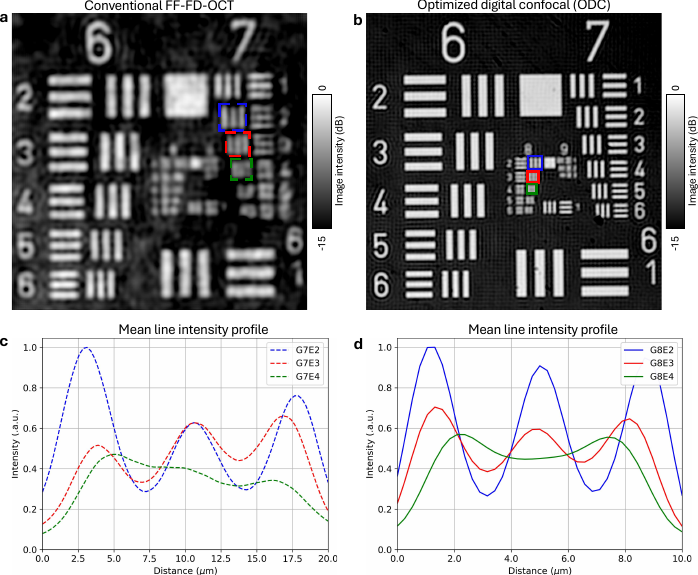}
\caption{USAF target showing doubling of the lateral resolution. \textbf{a, b.} Zoomed-in view of USAF target for conventional FF-FD-OCT and optimized digital confocal (ODC) enface, respectively. \textbf{c, d.} Average horizontal intensity in boxes shown in \textbf{a.} and \textbf{b.}, respectively.}
\label{fig:USAF}
\end{figure}

Lastly, we demonstrated improvement in lateral resolution in a USAF target. Figs.~\ref{fig:USAF}a and b show a zoomed-in view of the conventional FF-FD-OCT enface and the ODC enface, respectively. Here, the normalization of Eq.~\eqref{eq:est_eta} was not applied so that higher frequencies were not suppressed, allowing the resolution improvement to be assessed. Figs.~\ref{fig:USAF}c and d show the mean horizontal line intensity profiles of the boxes marked in Figs.~\ref{fig:USAF}a and b, corresponding to Elements 2, 3 and 4 for Group 7 and Group 8, respectively. The theoretical resolution of the system is given by $\frac{\lambda}{4\text{NA}} = 1.54\,\text{\textmu m}$, corresponding to Group 8 Element 3 (1.55\,\text{\textmu m}), which is indeed the smallest element that can still be resolved in the ODC enface. This is the expected doubling of the lateral resolution compared to using conventional FF-FD-OCT, where the smallest resolvable element is Group 7 Element 3 ($3.10\,\text{\textmu m}$).

Processing was performed on a computer with 128GB of RAM and an NVIDIA RTX 4090 (24GB VRAM). Initial reconstruction, $\vect{k}^{(i)}_\perp$-registration, shear correction and alignment of all volumes are performed in under 3 minutes. Subsequent local optimization can be performed efficiently by processing many isoplanatic patches in parallel on GPU. Optimization of the full digital confocal volume for the model eye (90 layers, 2304 isoplanatic patches of $32 \times 32$ pixels per layer, 20 iterations) took under 19 minutes.

\section{Discussion}
We demonstrated the effectiveness of the algorithm in various settings. The optimized digital confocal (ODC) images show substantial improvements over the conventional FF-FD-OCT images. For ODC image quality, correcting input aberrations proved as important as correcting output aberrations. In our setup, motion gives a random phase to each volume, which prevents coherent summing of the single scattering component in the resulting reconstructed images and is in principle indistinguishable from input aberrations. Considering the weak output aberrations in the lens tissue, the improvement from Fig.~\ref{fig:lens_tissue}a to b is almost entirely attributable to correction of input aberrations, yielding a 15 dB SNR gain. After the global correction of aberrations, volumes were mutually phase-stable. This is seen in the PSF phantom, where, after correcting the top layer, we find a clear depth-dependent defocus. This also indicates that the optimization finds physically consistent aberrations.

The ODC images showed a marked SNR improvement compared to the on-axis coherent averages (compare Figs.~\ref{fig:lens_tissue}d and e, Figs.~\ref{fig:model_eye}h and i, Figs.~\ref{fig:model_eye}j and k). This is unexpected since a phase-matched coherent average should provide the same SNR gain. We attribute the improvement to better rejection of autocorrelation and multiple-scattering noise. In on-axis coherent averaging, this noise adds constructively across phase-stable volumes and therefore scales similarly to the single-scattered signal. This leads to suboptimal improvement in SNR, which is theoretically only limited by Poisson-distributed shot noise. Under oblique illumination, however, the phase of multiply scattered and autocorrelation noise varies with the illumination, leading to partial cancellation during coherent summation. Consequently, oblique illumination can potentially suppress non-shot-noise contributions more effectively than conventional FF-FD-OCT. However, this advantage is lost when single-scattered signal is absent or very weak. This is observed in the model eye (Figs.~\ref{fig:model_eye}h, red inset), where local optimization increases the noise floor in the empty regions. This is caused by the algorithm attempting to maximize the coherent sum, without distinguishing signal from noise. This might be mitigated by introducing sparsity-promoting regularization, or by detecting speckle-like output aberrations and replacing them with estimates from neighboring patches.

Small features were much better visible in the ODC images. In the lens tissue, fibers were more clearly separated (Figs.~\ref{fig:lens_tissue}d and e), likely due to improved lateral resolution. In the model eye, many additional scatterers became visible (Figs.~\ref{fig:model_eye}h and i, white insets). Because these scatterers are comparable in size to those already visible in the on-axis image, it is unclear to what extent their improved visibility arises from enhanced resolution or increased SNR.

The least squares optimization is susceptible to convergence to local minima, maxima or saddle points, or a local gauge freedom in the reached point. In the PSF phantom, even after having corrected the full volume using the surface aberrations, aberrations at maximal depth could not be recovered without an initial estimate, instead converging to a speckle-like solution. The algorithm is thus not guaranteed to find the desired minimum when signal is weak or if the current solution is far away from this minimum. By working sequentially downwards using $\hat{P}_{\text{in/out}}(\vec{x}_\perp, z)$ as the initial estimate for $P_{\text{in/out},0}(\vec{x}_\perp, z+1)$, we remain close to the minimal loss value. This allowed convergence even at maximum depth.

Non-isoplanatic aberrations introduce additional minima when estimating the global pupil functions. This was observed in the model eye, where global estimation of the pupil functions was only successful when performed using the reflecting layer (marked in Fig.~\ref{fig:model_eye}j), which had most energy concentrated in a smaller number of isoplanatic patches.

Obtaining optimal image quality over the full FOV required a subsequent local optimization. The resolution of the aberration is determined by the size of the isoplanatic patch. Patch sizes must thus be large enough to capture the local PSF distortion, but small enough to allow a smooth transition from one patch to the next to minimize stitching artifacts. An initial global correction and volumetric refocusing are therefore advantageous, as they reduce PSF sidelobes and bring all patches closer to their local minimal loss values. However, small patches remain sensitive to noise and strong local reflections. The latter can be seen in Fig.~\ref{fig:lens_tissue}f, where a highly reflective fiber at the center produces local aberration estimates that unexpectedly deviate from neighboring patches.

Accurate $\vect{k}^{(i)}_\perp$ registration was essential for obtaining optimal image quality in the model eye. Notably, even small deviations from the theoretical values degraded the estimated output aberration. This will be even more important in \textit{in vivo} retinal imaging. Incorrect $\vect{k}^{(i)}_\perp$ values mix phase and amplitude from different parts of the object's $\vect{k}_\perp$-space when constructing $\tilde{\eta}_{j+1}(\vect{k}_\perp)$, preventing optimal coherent summation of the apertures. Because the object's phase is effectively random, even small errors strongly disturb its phase, particularly when considering larger patches due to the higher resolution of each pixel in $\vect{k}_\perp$-space. When subsequently estimating the pupil functions, the object's phase does not cancel when calculating the $P^*_{\text{in/out},j}\tilde{\eta}_{j+1}$ terms (Eqs.~\ref{eq:est_pin} and \ref{eq:est_pout}), which introduces phase noise.

Creating the oblique-illumination-capable FF-FD-OCT system required few hardware modifications, needing only a galvo and a synchronization device. However, some features of conventional FF-FD-OCT are lost. First, this setup is incompatible with multimode illumination due to its Mach-Zehnder configuration using a fiber coupler. In addition, the scanning changes the phase front of the illumination between volumes beyond the expected phase gradient since each collimated beam passes through different regions of the lenses. Furthermore, strong \textit{in vivo} sample motion will uniquely disturb the phases contained in each aperture, which will need to be addressed to make aberration correction feasible. Finally, the high volumetric rate that allows FF-FD-OCT to capture dynamic changes, such as optoretinography \cite{hillmann_vivo_2016} or dynamic OCT \cite{apelian_dynamic_2016}, is lost to a certain extent.

\section{Conclusion}
FF-FD-OCT offers a substantial speed advantage over point-scanning OCT but lacks a confocal gate. Previous work has demonstrated how oblique illumination may be used in full-field systems to synthesize a confocal volume in post-processing. We demonstrated such an oblique-illumination FF-FD-OCT system in a Mach-Zehnder configuration. Compared with previous work using a Michelson interferometer, our setup samples the object's spatial frequencies more efficiently, achieving an approximate fourfold improvement in SBP. Combined with a high-speed camera, this enabled volumetric acquisition at $18.3$ gigavoxels per second. An illumination-dependent shear introduced by the separation of the object and reference arms was corrected computationally. In addition, an accurate registration of the illumination wave vectors $\vect{k}^{(i)}_\perp$ was required in post-processing for optimal image formation.

We addressed aberrations by posing the scattering behavior as an overdetermined system, which could be leveraged to solve jointly for object, input aberrations and output aberrations, without requiring priors or assumptions. To this end, we formulated a closed-form solution of an alternating least squares problem, resulting in an iterative optimization algorithm. Notably, this algorithm is equivalent to the state-of-the-art CLASS algorithm, although it was derived without reference to the reflection matrix framework.

Using this algorithm, we reconstructed digital confocal volumes with globally, depth-dependent, and locally optimized focusing quality in various objects. The resulting images showed substantial improvements in resolution, contrast, and SNR. The reconstructed volumes also showed markedly higher SNR compared to identically corrected conventional FF-FD-OCT data due to better rejection of multiple-scattering and autocorrelation noise. However, the optimization was susceptible to local minima in regions with weak signal or strong anisoplanatism. Expansion of the scattering model or adding regularization may thus be required to improve robustness. In the future, our highly efficient acquisition of oblique illumination datasets at large FOV and high resolution may open up opportunities for applications in human \textit{in vivo} imaging.

\begin{backmatter}
\bmsection{Funding}
Dutch Research Council (NWO, OCENW.M.22.087 and OCENW.M.23.316). European Research Council (ERC, FunMicroRetina, grant agreement No. 101170453).

\bmsection{Acknowledgment}
We would like to thank ASML for lending us the high-speed camera, and Johannes F. de Boer for useful discussions.

\bmsection{Disclosures}
The authors declare no conflicts of interest.


\end{backmatter}

\bibliography{references}

\begin{thebibliography}{10}
\newcommand{\enquote}[1]{``#1''}

\bibitem{huang_optical_1991}
D.~Huang, E.~A. Swanson, C.~P. Lin, \emph{et~al.}, \enquote{Optical {Coherence} {Tomography},} {\protect\JournalTitle{Science}} \textbf{254}, 1178--1181 (1991).

\bibitem{boer_improved_2003}
J.~F.~d. Boer, B.~Cense, B.~H. Park, \emph{et~al.}, \enquote{Improved signal-to-noise ratio in spectral-domain compared with time-domain optical coherence tomography,} {\protect\JournalTitle{Optics Letters}} \textbf{28}, 2067--2069 (2003).

\bibitem{leitgeb_performance_2003}
R.~Leitgeb, C.~K. Hitzenberger, and A.~F. Fercher, \enquote{Performance of fourier domain vs. time domain optical coherence tomography,} {\protect\JournalTitle{Optics Express}} \textbf{11}, 889--894 (2003).

\bibitem{yun_high-speed_2003}
S.~H. Yun, G.~J. Tearney, J.~F.~d. Boer, \emph{et~al.}, \enquote{High-speed optical frequency-domain imaging,} {\protect\JournalTitle{Optics Express}} \textbf{11}, 2953--2963 (2003).

\bibitem{beaurepaire_full-field_1998}
E.~Beaurepaire, A.~C. Boccara, M.~Lebec, \emph{et~al.}, \enquote{Full-field optical coherence microscopy,} {\protect\JournalTitle{Optics Letters}} \textbf{23}, 244--246 (1998).

\bibitem{povazay_full-field_2006}
B.~Považay, A.~Unterhuber, B.~Hermann, \emph{et~al.}, \enquote{Full-field time-encoded frequency-domain optical coherence tomography,} {\protect\JournalTitle{Optics Express}} \textbf{14}, 7661--7669 (2006).

\bibitem{hillmann_aberration-free_2016}
D.~Hillmann, H.~Spahr, C.~Hain, \emph{et~al.}, \enquote{Aberration-free volumetric high-speed imaging of in vivo retina,} {\protect\JournalTitle{Scientific Reports}} \textbf{6}, 35209 (2016).

\bibitem{ntziachristos_going_2010}
V.~Ntziachristos, \enquote{Going deeper than microscopy: the optical imaging frontier in biology,} {\protect\JournalTitle{Nature Methods}} \textbf{7}, 603--614 (2010).

\bibitem{bertolotti_imaging_2022}
J.~Bertolotti and O.~Katz, \enquote{Imaging in complex media,} {\protect\JournalTitle{Nature Physics}} \textbf{18}, 1008--1017 (2022).

\bibitem{badon_multiple_2017}
A.~Badon, A.~C. Boccara, G.~Lerosey, \emph{et~al.}, \enquote{Multiple scattering limit in optical microscopy,} {\protect\JournalTitle{Optics Express}} \textbf{25}, 28914--28934 (2017).

\bibitem{booth_adaptive_2014}
M.~J. Booth, \enquote{Adaptive optical microscopy: the ongoing quest for a perfect image,} {\protect\JournalTitle{Light: Science \& Applications}} \textbf{3}, e165--e165 (2014).

\bibitem{ji_adaptive_2017}
N.~Ji, \enquote{Adaptive optical fluorescence microscopy,} {\protect\JournalTitle{Nature Methods}} \textbf{14}, 374--380 (2017).

\bibitem{park_perspective_2018}
J.-H. Park, Z.~Yu, K.~Lee, \emph{et~al.}, \enquote{Perspective: {Wavefront} shaping techniques for controlling multiple light scattering in biological tissues: {Toward} in vivo applications,} {\protect\JournalTitle{APL Photonics}} \textbf{3}, 100901 (2018).

\bibitem{kumar_numerical_2014}
A.~Kumar, W.~Drexler, and R.~A. Leitgeb, \enquote{Numerical focusing methods for full field {OCT}: a comparison based on a common signal model,} {\protect\JournalTitle{Optics Express}} \textbf{22}, 16061--16078 (2014).

\bibitem{ralston_interferometric_2007}
T.~S. Ralston, D.~L. Marks, P.~S. Carney, and S.~A. Boppart, \enquote{Interferometric synthetic aperture microscopy,} {\protect\JournalTitle{Nature physics}} \textbf{3}, 129--134 (2007).

\bibitem{shemonski_computational_2015}
N.~D. Shemonski, F.~A. South, Y.-Z. Liu, \emph{et~al.}, \enquote{Computational high-resolution optical imaging of the living human retina,} {\protect\JournalTitle{Nature Photonics}} \textbf{9}, 440--443 (2015).

\bibitem{adie_computational_2012}
S.~G. Adie, B.~W. Graf, A.~Ahmad, \emph{et~al.}, \enquote{Computational adaptive optics for broadband optical interferometric tomography of biological tissue,} {\protect\JournalTitle{Proceedings of the National Academy of Sciences}} \textbf{109}, 7175--7180 (2012).

\bibitem{adie_guide-star-based_2012}
S.~G. Adie, N.~D. Shemonski, B.~W. Graf, \emph{et~al.}, \enquote{Guide-star-based computational adaptive optics for broadband interferometric tomography,} {\protect\JournalTitle{Applied Physics Letters}} \textbf{101}, 221117 (2012).

\bibitem{liu_phase_2013}
C.~Liu, X.~Yu, and M.~K. Kim, \enquote{Phase aberration correction by correlation in digital holographic adaptive optics,} {\protect\JournalTitle{Applied Optics}} \textbf{52}, 2940--2949 (2013).

\bibitem{thurman_phase-error_2008}
S.~T. Thurman and J.~R. Fienup, \enquote{Phase-error correction in digital holography,} {\protect\JournalTitle{JOSA A}} \textbf{25}, 983--994 (2008).

\bibitem{ginner_noniterative_2017}
L.~Ginner, A.~Kumar, D.~Fechtig, \emph{et~al.}, \enquote{Noniterative digital aberration correction for cellular resolution retinal optical coherence tomography in vivo,} {\protect\JournalTitle{Optica}} \textbf{4}, 924--931 (2017).

\bibitem{kumar_subaperture_2013}
A.~Kumar, W.~Drexler, and R.~A. Leitgeb, \enquote{Subaperture correlation based digital adaptive optics for full field optical coherence tomography,} {\protect\JournalTitle{Optics Express}} \textbf{21}, 10850--10866 (2013).

\bibitem{hillmann_computational_2019}
D.~Hillmann, C.~Pfäffle, H.~Spahr, \emph{et~al.}, \enquote{Computational adaptive optics for optical coherence tomography using multiple randomized subaperture correlations,} {\protect\JournalTitle{Optics Letters}} \textbf{44}, 3905--3908 (2019).

\bibitem{aubry_detection_2009}
A.~Aubry and A.~Derode, \enquote{Detection and imaging in a random medium: {A} matrix method to overcome multiple scattering and aberration,} {\protect\JournalTitle{Journal of Applied Physics}} \textbf{106}, 044903 (2009).

\bibitem{lambert_reflection_2020}
W.~Lambert, \enquote{Reflection {Matrix} {Approach} for {Quantitative} {Imaging} of {Scattering} {Media},} {\protect\JournalTitle{Physical Review X}} \textbf{10} (2020).

\bibitem{popoff_controlling_2011}
S.~M. Popoff, G.~Lerosey, M.~Fink, \emph{et~al.}, \enquote{Controlling light through optical disordered media: transmission matrix approach,} {\protect\JournalTitle{New Journal of Physics}} \textbf{13}, 123021 (2011).

\bibitem{badon_smart_2016}
A.~Badon, D.~Li, G.~Lerosey, \emph{et~al.}, \enquote{Smart optical coherence tomography for ultra-deep imaging through highly scattering media,} {\protect\JournalTitle{Science Advances}} \textbf{2}, e1600370 (2016).

\bibitem{kang_high-resolution_2017}
S.~Kang, P.~Kang, S.~Jeong, \emph{et~al.}, \enquote{High-resolution adaptive optical imaging within thick scattering media using closed-loop accumulation of single scattering,} {\protect\JournalTitle{Nature Communications}} \textbf{8}, 2157 (2017).

\bibitem{yoon_laser_2020}
S.~Yoon, H.~Lee, J.~H. Hong, \emph{et~al.}, \enquote{Laser scanning reflection-matrix microscopy for aberration-free imaging through intact mouse skull,} {\protect\JournalTitle{Nature Communications}} \textbf{11}, 5721 (2020).

\bibitem{najar_harnessing_2024}
U.~Najar, V.~Barolle, P.~Balondrade, \emph{et~al.}, \enquote{Harnessing forward multiple scattering for optical imaging deep inside an opaque medium,} {\protect\JournalTitle{Nature Communications}} \textbf{15}, 7349 (2024).

\bibitem{balondrade_multi-spectral_2024}
P.~Balondrade, V.~Barolle, N.~Guigui, \emph{et~al.}, \enquote{Multi-spectral reflection matrix for ultrafast {3D} label-free microscopy,} {\protect\JournalTitle{Nature Photonics}} \textbf{18}, 1097--1104 (2024).

\bibitem{giraudat_label-free_2026}
E.~Giraudat, V.~Barolle, F.~Bureau, \emph{et~al.}, \enquote{Label-free subcellular {3D} imaging of oocytes and embryos via reflection matrix microscopy,}  (2026). ArXiv:2410.11126 [physics.optics].

\bibitem{kim_label-free_2019}
M.~Kim, Y.~Jo, J.~H. Hong, \emph{et~al.}, \enquote{Label-free neuroimaging in vivo using synchronous angular scanning microscopy with single-scattering accumulation algorithm,} {\protect\JournalTitle{Nature Communications}} \textbf{10}, 3152 (2019).

\bibitem{lohmann_spacebandwidth_1996}
A.~W. Lohmann, R.~G. Dorsch, D.~Mendlovic, \emph{et~al.}, \enquote{Space–bandwidth product of optical signals and systems,} {\protect\JournalTitle{JOSA A}} \textbf{13}, 470--473 (1996).

\bibitem{park_review_2021}
J.~Park, D.~J. Brady, G.~Zheng, \emph{et~al.}, \enquote{Review of bio-optical imaging systems with a high space-bandwidth product,} {\protect\JournalTitle{Advanced Photonics}} \textbf{3}, 044001 (2021).

\bibitem{huang_optimization_2014}
X.~Huang, H.~Yan, R.~Harder, \emph{et~al.}, \enquote{Optimization of overlap uniformness for ptychography,} {\protect\JournalTitle{Optics Express}} \textbf{22}, 12634--12644 (2014).

\bibitem{heintzmann_super-resolution_2017}
R.~Heintzmann and T.~Huser, \enquote{Super-{Resolution} {Structured} {Illumination} {Microscopy},} {\protect\JournalTitle{Chemical Reviews}} \textbf{117}, 13890--13908 (2017).

\bibitem{lambert_ultrasound_2022}
W.~Lambert, L.~A. Cobus, J.~Robin, \emph{et~al.}, \enquote{Ultrasound {Matrix} {Imaging}—{Part} {II}: {The} {Distortion} {Matrix} for {Aberration} {Correction} {Over} {Multiple} {Isoplanatic} {Patches},} {\protect\JournalTitle{IEEE Transactions on Medical Imaging}} \textbf{41}, 3921--3938 (2022).

\bibitem{thrun_graph_2006}
S.~Thrun and M.~Montemerlo, \enquote{The {Graph} {SLAM} {Algorithm} with {Applications} to {Large}-{Scale} {Mapping} of {Urban} {Structures},} {\protect\JournalTitle{The International Journal of Robotics Research}} \textbf{25}, 403--429 (2006).

\bibitem{kang_imaging_2015}
S.~Kang, S.~Jeong, W.~Choi, \emph{et~al.}, \enquote{Imaging deep within a scattering medium using collective accumulation of single-scattered waves,} {\protect\JournalTitle{Nature Photonics}} \textbf{9}, 253--258 (2015).

\bibitem{kwon_computational_2023}
Y.~Kwon, J.~H. Hong, S.~Kang, \emph{et~al.}, \enquote{Computational conjugate adaptive optics microscopy for longitudinal through-skull imaging of cortical myelin,} {\protect\JournalTitle{Nature Communications}} \textbf{14}, 105 (2023).

\bibitem{hillmann_vivo_2016}
D.~Hillmann, H.~Spahr, C.~Pfäffle, \emph{et~al.}, \enquote{In vivo optical imaging of physiological responses to photostimulation in human photoreceptors,} {\protect\JournalTitle{Proceedings of the National Academy of Sciences}} \textbf{113}, 13138--13143 (2016).

\bibitem{apelian_dynamic_2016}
C.~Apelian, F.~Harms, O.~Thouvenin, and A.~C. Boccara, \enquote{Dynamic full field optical coherence tomography: subcellular metabolic contrast revealed in tissues by interferometric signals temporal analysis,} {\protect\JournalTitle{Biomedical Optics Express}} \textbf{7}, 1511--1524 (2016).

\end{thebibliography}

\end{document}